%Paper: hep-th/9212075
%From: christof@theory3.caltech.edu (Christof Schmidhuber)
%Date: Thu, 10 Dec 92 23:12:41 PST
%Date (revised): Fri, 11 Dec 92 23:59:55 PST
%Date (revised): Tue, 16 Feb 93 20:37:33 PST

%     This is setup.tex and phyzzx.tex ,merged , and with
%     hoffset and voffset changed to suit the Apple laserwriter
%     November 88.
%%%%%%%%%%%%%%%%%%%%%%%%%%%%%%%%%%%%%%%%%%%%%%%%%%%%%%%%%%%%%%%%%%%%%%%%%
% % % % % % % % % % % % % % % % % % % % % % % % % % % % % % % % % % % %
%%%   This is PHYZZX macro package.   % % % % % % % % % % % % % % % % %
%% % % % % % % % % % % % % % % % % % % % % % % % % % % % % % % % % % % %
%%%  This version of PHYZZX should be used with Version 1.0 of TEX  % %
%% % % % % % % % % % % % % % % % % % % % % % % % % % % % % % % % % % % %
%%%   Do not "\input phyzzx" unless you preload or "\input" PLAIN.  % %
%% % % % % % % % % % % % % % % % % % % % % % % % % % % % % % % % % % % %
%%%   To preload both PLAIN and PHYZZX, begin your file with    % % % %
%%%  a line "%macropackage=phyzzx" instead of "\input phyzzx".  % % % %
%% % % % % % % % % % % % % % % % % % % % % % % % % % % % % % % % % % % %
%%%%%%%%%%%%%%%%%%%%%%%%%%%%%%%%%%%%%%%%%%%%%%%%%%%%%%%%%%%%%%%%%%%%%%%%
%%%%%%%  Created by Vadim Kaplunovsky in June 1984.   %%%%%%%%%%%%%%%%%%
% % % % % % % % % % % % % % % % % % % % % % % % % % % % % % % % % % % %
%%%%%%%%%%%%  Latest update/debug: May 29, 1985   %%%%%%%%%%%%%%%%%%%%%%
%%%%%%%%%%%%%%%%%%%%%%%%%%%%%%%%%%%%%%%%%%%%%%%%%%%%%%%%%%%%%%%%%%%%%%%%
%
\expandafter\ifx\csname phyzzx\endcsname\relax\else
 \errhelp{Hit <CR> and go ahead.}
 \errmessage{PHYZZX macros are already loaded or input. }
 \endinput \fi
\catcode`\@=11 % This allows us to modify PLAIN macros.
%
%%%%%%%%%%%%%%%%%%%%%%%%%%%%%%%%%%%%%%%%%%%%%%%%%%%%%%%%%%%%%%%%%%%%%%%%
%
%   I begin with fonts.
%
\font\seventeenrm=cmr17
\font\fourteenrm=cmr12 scaled\magstep1
\font\twelverm=cmr12
\font\ninerm=cmr9            \font\sixrm=cmr6
%
%\font\seventeenbf=cmbx10 scaled\magstep3
\font\fourteenbf=cmbx10 scaled\magstep2
\font\twelvebf=cmbx12
\font\ninebf=cmbx9            \font\sixbf=cmbx6
%
%\font\seventeeni=cmmi10 scaled\magstep3     \skewchar\seventeeni='177
\font\fourteeni=cmmi10 scaled\magstep2      \skewchar\fourteeni='177
\font\twelvei=cmmi12			        \skewchar\twelvei='177
\font\ninei=cmmi9                           \skewchar\ninei='177
\font\sixi=cmmi6                            \skewchar\sixi='177
%
%\font\seventeensy=cmsy10 scaled\magstep3    \skewchar\seventeensy='60
\font\fourteensy=cmsy10 scaled\magstep2     \skewchar\fourteensy='60
\font\twelvesy=cmsy10 scaled\magstep1	    \skewchar\twelvesy='60
\font\ninesy=cmsy9                          \skewchar\ninesy='60
\font\sixsy=cmsy6                           \skewchar\sixsy='60
%
%\font\seventeenex=cmex10 scaled\magstep3
\font\fourteenex=cmex10 scaled\magstep2
\font\twelveex=cmex10 scaled\magstep1
%\font\elevenex=cmex10 scaled\magstephalf
%
%\font\seventeensl=cmsl10 scaled\magstep3
\font\fourteensl=cmsl12 scaled\magstep1
\font\twelvesl=cmsl12
\font\ninesl=cmsl9
%
%\font\seventeenit=cmti10 scaled\magstep3
\font\fourteenit=cmti12 scaled\magstep1
\font\twelveit=cmti12
\font\nineit=cmti9
\font\fourteentt=cmtt10 scaled\magstep2
\font\twelvett=cmtt12
\font\fourteencp=cmcsc10 scaled\magstep2
\font\twelvecp=cmcsc10 scaled\magstep1
\font\tencp=cmcsc10
\newfam\cpfam
\newdimen\b@gheight		\b@gheight=12pt
\newcount\f@ntkey		\f@ntkey=0
\def\f@m{\afterassignment\samef@nt\f@ntkey=}
\def\samef@nt{\fam=\f@ntkey \the\textfont\f@ntkey\relax}
\def\rm{\f@m0 }
\def\mit{\f@m1 }         
\def\cal{\f@m2 }
\def\it{\f@m\itfam}
\def\sl{\f@m\slfam}
\def\bf{\f@m\bffam}
\def\tt{\f@m\ttfam}
\def\caps{\f@m\cpfam}
\def\fourteenpoint{\relax
    \textfont0=\fourteenrm          \scriptfont0=\tenrm
      \scriptscriptfont0=\sevenrm
    \textfont1=\fourteeni           \scriptfont1=\teni
      \scriptscriptfont1=\seveni
    \textfont2=\fourteensy          \scriptfont2=\tensy
      \scriptscriptfont2=\sevensy
    \textfont3=\fourteenex          \scriptfont3=\twelveex
      \scriptscriptfont3=\tenex
    \textfont\itfam=\fourteenit     \scriptfont\itfam=\tenit
    \textfont\slfam=\fourteensl     \scriptfont\slfam=\tensl
    \textfont\bffam=\fourteenbf     \scriptfont\bffam=\tenbf
      \scriptscriptfont\bffam=\sevenbf
    \textfont\ttfam=\fourteentt
    \textfont\cpfam=\fourteencp
    \samef@nt
    \b@gheight=14pt
    \setbox\strutbox=\hbox{\vrule height 0.85\b@gheight
				depth 0.35\b@gheight width\z@ }}
\def\twelvepoint{\relax
    \textfont0=\twelverm          \scriptfont0=\ninerm
      \scriptscriptfont0=\sixrm
    \textfont1=\twelvei           \scriptfont1=\ninei
      \scriptscriptfont1=\sixi
    \textfont2=\twelvesy           \scriptfont2=\ninesy
      \scriptscriptfont2=\sixsy
    \textfont3=\twelveex          \scriptfont3=\tenex
      \scriptscriptfont3=\tenex
    \textfont\itfam=\twelveit     \scriptfont\itfam=\nineit
    \textfont\slfam=\twelvesl     \scriptfont\slfam=\ninesl
    \textfont\bffam=\twelvebf     \scriptfont\bffam=\ninebf
      \scriptscriptfont\bffam=\sixbf
    \textfont\ttfam=\twelvett
    \textfont\cpfam=\twelvecp
    \samef@nt
    \b@gheight=12pt
    \setbox\strutbox=\hbox{\vrule height 0.85\b@gheight
				depth 0.35\b@gheight width\z@ }}
\def\tenpoint{\relax
    \textfont0=\tenrm          \scriptfont0=\sevenrm
      \scriptscriptfont0=\fiverm
    \textfont1=\teni           \scriptfont1=\seveni
      \scriptscriptfont1=\fivei
    \textfont2=\tensy          \scriptfont2=\sevensy
      \scriptscriptfont2=\fivesy
    \textfont3=\tenex          \scriptfont3=\tenex
      \scriptscriptfont3=\tenex
    \textfont\itfam=\tenit     \scriptfont\itfam=\seveni
    \textfont\slfam=\tensl     \scriptfont\slfam=\sevenrm
    \textfont\bffam=\tenbf     \scriptfont\bffam=\sevenbf
      \scriptscriptfont\bffam=\fivebf
    \textfont\ttfam=\tentt
    \textfont\cpfam=\tencp
    \samef@nt
    \b@gheight=10pt
    \setbox\strutbox=\hbox{\vrule height 0.85\b@gheight
				depth 0.35\b@gheight width\z@ }}
%
%%%%%%%%%%%%%%%%%%%%%%%%%%%%%%%%%%%%%%%%%%%%%%%%%%%%%%%%%%%%%%%%%%%%%%%%
%
%   Next, I define basic spacing parameters.
%
\normalbaselineskip = 20pt plus 0.2pt minus 0.1pt
\normallineskip = 1.5pt plus 0.1pt minus 0.1pt
\normallineskiplimit = 1.5pt
\newskip\normaldisplayskip
\normaldisplayskip = 20pt plus 5pt minus 10pt
\newskip\normaldispshortskip
\normaldispshortskip = 6pt plus 5pt
\newskip\normalparskip
\normalparskip = 6pt plus 2pt minus 1pt
\newskip\skipregister
\skipregister = 5pt plus 2pt minus 1.5pt
\newif\ifsingl@    \newif\ifdoubl@
\newif\iftwelv@    \twelv@true
\def\singlespace{\singl@true\doubl@false\spaces@t}
\def\doublespace{\singl@false\doubl@true\spaces@t}
\def\normalspace{\singl@false\doubl@false\spaces@t}
\def\Tenpoint{\tenpoint\twelv@false\spaces@t}
\def\Twelvepoint{\twelvepoint\twelv@true\spaces@t}
\def\spaces@t{\relax
      \iftwelv@ \ifsingl@\subspaces@t3:4;\else\subspaces@t1:1;\fi
       \else \ifsingl@\subspaces@t3:5;\else\subspaces@t4:5;\fi \fi
      \ifdoubl@ \multiply\baselineskip by 5
         \divide\baselineskip by 4 \fi }
\def\subspaces@t#1:#2;{
      \baselineskip = \normalbaselineskip
      \multiply\baselineskip by #1 \divide\baselineskip by #2
      \lineskip = \normallineskip
      \multiply\lineskip by #1 \divide\lineskip by #2
      \lineskiplimit = \normallineskiplimit
      \multiply\lineskiplimit by #1 \divide\lineskiplimit by #2
      \parskip = \normalparskip
      \multiply\parskip by #1 \divide\parskip by #2
      \abovedisplayskip = \normaldisplayskip
      \multiply\abovedisplayskip by #1 \divide\abovedisplayskip by #2
      \belowdisplayskip = \abovedisplayskip
      \abovedisplayshortskip = \normaldispshortskip
      \multiply\abovedisplayshortskip by #1
        \divide\abovedisplayshortskip by #2
      \belowdisplayshortskip = \abovedisplayshortskip
      \advance\belowdisplayshortskip by \belowdisplayskip
      \divide\belowdisplayshortskip by 2
      \smallskipamount = \skipregister
      \multiply\smallskipamount by #1 \divide\smallskipamount by #2
      \medskipamount = \smallskipamount \multiply\medskipamount by 2
      \bigskipamount = \smallskipamount \multiply\bigskipamount by 4 }
\def\normalbaselines{ \baselineskip=\normalbaselineskip
   \lineskip=\normallineskip \lineskiplimit=\normallineskip
   \iftwelv@\else \multiply\baselineskip by 4 \divide\baselineskip by 5
     \multiply\lineskiplimit by 4 \divide\lineskiplimit by 5
     \multiply\lineskip by 4 \divide\lineskip by 5 \fi }
\Twelvepoint  % That's the default
\interlinepenalty=50
\interfootnotelinepenalty=5000
\predisplaypenalty=9000
\postdisplaypenalty=500
\hfuzz=1pt
\vfuzz=0.2pt
\voffset=0pt
\dimen\footins=8 truein
%
%%%%%%%%%%%%%%%%%%%%%%%%%%%%%%%%%%%%%%%%%%%%%%%%%%%%%%%%%%%%%%%%%%%%%%%%
%
%   Next, I define output routines, footnotes & related stuff.
%
\def\pagecontents{
   \ifvoid\topins\else\unvbox\topins\vskip\skip\topins\fi
   \dimen@ = \dp255 \unvbox255
   \ifvoid\footins\else\vskip\skip\footins\footrule\unvbox\footins\fi
   \ifr@ggedbottom \kern-\dimen@ \vfil \fi }
\def\makeheadline{\vbox to 0pt{ \skip@=\topskip
      \advance\skip@ by -12pt \advance\skip@ by -2\normalbaselineskip
      \vskip\skip@ \line{\vbox to 12pt{}\the\headline} \vss
      }\nointerlineskip}
\def\makefootline{\baselineskip = 1.5\normalbaselineskip
                 \line{\the\footline}}
\newif\iffrontpage
\newif\ifletterstyle
\newif\ifp@genum
\def\nopagenumbers{\p@genumfalse}
\def\pagenumbers{\p@genumtrue}
\pagenumbers
\newtoks\paperheadline
\newtoks\letterheadline
\newtoks\paperfootline
\newtoks\letterfootline
\newtoks\letterinfo
\newtoks\Letterinfo
\newtoks\date
\footline={\ifletterstyle\the\letterfootline\else\the\paperfootline\fi}
\paperfootline={\hss\iffrontpage\else\ifp@genum\tenrm\folio\hss\fi\fi}
\letterfootline={\iffrontpage\LETTERFOOT\else\hfil\fi}
\Letterinfo={\hfil}
\letterinfo={\hfil}
\def\LETTERFOOT{\hfil} %\ninerm PASADENA, CALIFORNIA 91125 TELEPHONE:
%(818)356-6686\hfil}
%
\def\LETTERHEAD{\vtop{\baselineskip=20pt\hbox to
\hsize{\hfil\seventeenrm\strut
CALIFORNIA INSTITUTE OF TECHNOLOGY \hfil}
\hbox to \hsize{\hfil\ninerm\strut
CHARLES C. LAURITSEN LABORATORY OF HIGH ENERGY PHYSICS \hfil}
\hbox to \hsize{\hfil\ninerm\strut
PASADENA, CALIFORNIA 91125 \hfil}}}
\headline={\ifletterstyle\the\letterheadline\else\the\paperheadline\fi}
\paperheadline={\hfil}
\letterheadline{\iffrontpage \LETTERHEAD\else
    \rm \ifp@genum \hfil \folio\hfil\fi\fi}
\def\monthname{\relax\ifcase\month 0/\or January\or February\or
   March\or April\or May\or June\or July\or August\or September\or
   October\or November\or December\else\number\month/\fi}
\def\today{\monthname\ \number\day, \number\year}
\date={\today}
\countdef\pageno=1      \countdef\pagen@=0
\countdef\pagenumber=1  \pagenumber=1
\def\advancepageno{\global\advance\pagen@ by 1
   \ifnum\pagenumber<0 \global\advance\pagenumber by -1
    \else\global\advance\pagenumber by 1 \fi \global\frontpagefalse }
\def\folio{\ifnum\pagenumber<0 \romannumeral-\pagenumber
           \else \number\pagenumber \fi }
\def\footrule{\dimen@=\prevdepth\nointerlineskip
   \vbox to 0pt{\vskip -0.25\baselineskip \hrule width 0.35\hsize \vss}
   \prevdepth=\dimen@ }
\newtoks\foottokens
\foottokens={}
\newdimen\footindent
\footindent=24pt
\def\vfootnote#1{\insert\footins\bgroup
   \interlinepenalty=\interfootnotelinepenalty \floatingpenalty=20000
   \singl@true\doubl@false\Tenpoint
   \splittopskip=\ht\strutbox \boxmaxdepth=\dp\strutbox
   \leftskip=\footindent \rightskip=\z@skip
   \parindent=0.5\footindent \parfillskip=0pt plus 1fil
   \spaceskip=\z@skip \xspaceskip=\z@skip
   \the\foottokens
   \Textindent{$ #1 $}\footstrut\futurelet\next\fo@t}
\def\Textindent#1{\noindent\llap{#1\enspace}\ignorespaces}
\def\footnote#1{\attach{#1}\vfootnote{#1}}

\let\footsymbol=\star
\newcount\lastf@@t           \lastf@@t=-1
\newcount\footsymbolcount    \footsymbolcount=0
\newif\ifPhysRev
\def\bumpfootsymbolcount{\relax
   \iffrontpage \bumpfootsymbolNP \else \advance\lastf@@t by 1
     \ifPhysRev \bumpfootsymbolPR \else \bumpfootsymbolNP \fi \fi
   \global\lastf@@t=\pagen@ }
\def\bumpfootsymbolNP{\ifnum\footsymbolcount <0 \global\footsymbolcount =0 \fi
    \ifnum\lastf@@t<\pagen@ \global\footsymbolcount=0
     \else \global\advance\footsymbolcount by 1 \fi }
\def\bumpfootsymbolPR{\ifnum\footsymbolcount >0 \global\footsymbolcount =0 \fi
      \global\advance\footsymbolcount by -1 }
\def\fd@f#1 {\xdef\footsymbol{\mathchar"#1 }}
\def\generatefootsymbol{\ifcase\footsymbolcount \fd@f 13F \or \fd@f 279
	\or \fd@f 27A \or \fd@f 278 \or \fd@f 27B \else
	\ifnum\footsymbolcount <0 \fd@f{023 \number-\footsymbolcount }
	 \else \fd@f 203 {\loop \ifnum\footsymbolcount >5
		\fd@f{203 \footsymbol } \advance\footsymbolcount by -1
		\repeat }\fi \fi }

\def\nonfrenchspacing{\sfcode`\.=3001 \sfcode`\!=3000 \sfcode`\?=3000
	\sfcode`\:=2000 \sfcode`\;=1500 \sfcode`\,=1251 }
\nonfrenchspacing
\newdimen\d@twidth
 {\setbox0=\hbox{s.} \global\d@twidth=\wd0 \setbox0=\hbox{s}
	\global\advance\d@twidth by -\wd0 }
\def\removehglue{\loop \unskip \ifdim\lastskip >\z@ \repeat }
\def\roll@ver#1{\removehglue \nobreak \count255 =\spacefactor \dimen@=\z@
	\ifnum\count255 =3001 \dimen@=\d@twidth \fi
	\ifnum\count255 =1251 \dimen@=\d@twidth \fi
    \iftwelv@ \kern-\dimen@ \else \kern-0.83\dimen@ \fi
   #1\spacefactor=\count255 }
\def\step@ver#1{\relax \ifmmode #1\else \ifhmode
	\roll@ver{${}#1$}\else {\setbox0=\hbox{${}#1$}}\fi\fi }
\def\attach#1{\step@ver{\strut^{\mkern 2mu #1} }}
%
%%%%%%%%%%%%%%%%%%%%%%%%%%%%%%%%%%%%%%%%%%%%%%%%%%%%%%%%%%%%%%%%%%%%%%%%
%
%   Here come chapter, section, subsection & appendix macros.
%
\newcount\chapternumber      \chapternumber=0
\newcount\sectionnumber      \sectionnumber=0
\newcount\equanumber         \equanumber=0
\let\chapterlabel=\relax
\let\sectionlabel=\relax
\newtoks\chapterstyle        \chapterstyle={\Number}
\newtoks\sectionstyle        \sectionstyle={\chapterlabel\Number}
\newskip\chapterskip         \chapterskip=\bigskipamount
\newskip\sectionskip         \sectionskip=\medskipamount
\newskip\headskip            \headskip=8pt plus 3pt minus 3pt
\newdimen\chapterminspace    \chapterminspace=15pc
\newdimen\sectionminspace    \sectionminspace=10pc
\newdimen\referenceminspace  \referenceminspace=25pc
\def\chapterreset{\global\advance\chapternumber by 1
   \ifnum\equanumber<0 \else\global\equanumber=0\fi
   \sectionnumber=0 \makechapterlabel}
\def\makechapterlabel{\let\sectionlabel=\relax
   \xdef\chapterlabel{\the\chapterstyle{\the\chapternumber}.}}
\def\alphabetic#1{\count255='140 \advance\count255 by #1\char\count255}
\def\Alphabetic#1{\count255='100 \advance\count255 by #1\char\count255}
\def\Roman#1{\uppercase\expandafter{\romannumeral #1}}
\def\roman#1{\romannumeral #1}
\def\Number#1{\number #1}
\def\BLANC#1{}
\def\titlestyle#1{\par\begingroup \interlinepenalty=9999
     \leftskip=0.02\hsize plus 0.23\hsize minus 0.02\hsize
     \rightskip=\leftskip \parfillskip=0pt
     \hyphenpenalty=9000 \exhyphenpenalty=9000
     \tolerance=9999 \pretolerance=9000
     \spaceskip=0.333em \xspaceskip=0.5em
     \iftwelv@\fourteenpoint\else\twelvepoint\fi
   \noindent #1\par\endgroup }
\def\spacecheck#1{\dimen@=\pagegoal\advance\dimen@ by -\pagetotal
   \ifdim\dimen@<#1 \ifdim\dimen@>0pt \vfil\break \fi\fi}
\def\TableOfContentEntry#1#2#3{\relax}
\def\chapter#1{\par \penalty-300 \vskip\chapterskip
   \spacecheck\chapterminspace
   \chapterreset \titlestyle{\chapterlabel\ #1}
   \TableOfContentEntry c\chapterlabel{#1}
   \nobreak\vskip\headskip \penalty 30000
   \wlog{\string\chapter\space \chapterlabel} }

\def\section#1{\par \ifnum\the\lastpenalty=30000\else
   \penalty-200\vskip\sectionskip \spacecheck\sectionminspace\fi
   \global\advance\sectionnumber by 1
   \xdef\sectionlabel{\the\sectionstyle\the\sectionnumber}
   \wlog{\string\section\space \sectionlabel}
   \TableOfContentEntry s\sectionlabel{#1}
   \noindent {\caps\enspace\sectionlabel\quad #1}\par
   \nobreak\vskip\headskip \penalty 30000 }
\def\subsection#1{\par
   \ifnum\the\lastpenalty=30000\else \penalty-100\smallskip \fi
   \noindent\undertext{#1}\enspace \vadjust{\penalty5000}}

\def\undertext#1{\vtop{\hbox{#1}\kern 1pt \hrule}}
\def\APPENDIX#1#2{\par\penalty-300\vskip\chapterskip
   \spacecheck\chapterminspace \chapterreset \xdef\chapterlabel{#1}
   \titlestyle{APPENDIX #2} \nobreak\vskip\headskip \penalty 30000
   \TableOfContentEntry a{#1}{#2}
   \wlog{\string\Appendix\ \chapterlabel} }
\def\Appendix#1{\APPENDIX{#1}{#1}}
\def\appendix{\APPENDIX{A}{}}
\def\unnumberedchapters{\let\makechapterlabel=\relax \let\chapterlabel=\relax
   \sectionstyle={\BLANC}\let\sectionlabel=\relax \sequentialequations }
%
%%%%%%%%%%%%%%%%%%%%%%%%%%%%%%%%%%%%%%%%%%%%%%%%%%%%%%%%%%%%%%%%%%%%%%%%
%
%   Here come macros for equation numbering.
%
\def\eqname#1{\relax \ifnum\equanumber<0
     \xdef#1{{\noexpand\rm(\number-\equanumber)}}%
       \global\advance\equanumber by -1
    \else \global\advance\equanumber by 1
      \xdef#1{{\noexpand\rm(\chapterlabel\number\equanumber)}} \fi #1}
\def\eqinsert#1{\noalign{\dimen@=\prevdepth \nointerlineskip
   \setbox0=\hbox to\displaywidth{\hfil #1}
   \vbox to 0pt{\kern 0.5\baselineskip\hbox{$\!\box0\!$}\vss}
   \prevdepth=\dimen@}}
%

%
%%%%%%%%%%%%%%%%%%%%%%%%%%%%%%%%%%%%%%%%%%%%%%%%%%%%%%%%%%%%%%%%%%%%%%%%
%   Here come items and lists
%
\def\GENITEM#1;#2{\par \hangafter=0 \hangindent=#1
    \Textindent{$ #2 $}\ignorespaces}
\outer\def\newitem#1=#2;{\gdef#1{\GENITEM #2;}}
\newdimen\itemsize                \itemsize=30pt
\newitem\item=1\itemsize;
\newitem\sitem=1.75\itemsize;     
\newitem\ssitem=2.5\itemsize;     
\outer\def\newlist#1=#2&#3&#4;{\toks0={#2}\toks1={#3}%
   \count255=\escapechar \escapechar=-1
   \alloc@0\list\countdef\insc@unt\listcount     \listcount=0
   \edef#1{\par
      \countdef\listcount=\the\allocationnumber
      \advance\listcount by 1
      \hangafter=0 \hangindent=#4
      \Textindent{\the\toks0{\listcount}\the\toks1}}
   \expandafter\expandafter\expandafter
    \edef\c@t#1{begin}{\par
      \countdef\listcount=\the\allocationnumber \listcount=1
      \hangafter=0 \hangindent=#4
      \Textindent{\the\toks0{\listcount}\the\toks1}}
   \expandafter\expandafter\expandafter
    \edef\c@t#1{con}{\par \hangafter=0 \hangindent=#4 \noindent}
   \escapechar=\count255}
\def\c@t#1#2{\csname\string#1#2\endcsname}
\newlist\point=\Number&.&1.0\itemsize;
\newlist\subpoint=(\alphabetic&)&1.75\itemsize;
\newlist\subsubpoint=(\roman&)&2.5\itemsize;
%

%
%%%%%%%%%%%%%%%%%%%%%%%%%%%%%%%%%%%%%%%%%%%%%%%%%%%%%%%%%%%%%%%%%%%%%%%%
%
%   Here come macros for references, figures & tables.
%
% % % % % % % % % % % % % % % % % % % % % % % % % % % % % % % % % % % %
%%  First, references.
%
\newcount\referencecount     \referencecount=0
\newcount\lastrefsbegincount \lastrefsbegincount=0
\newif\ifreferenceopen       \newwrite\referencewrite
\newif\ifrw@trailer
\newdimen\refindent     \refindent=30pt
\def\NPrefmark#1{\attach{\scriptscriptstyle [ #1 ] }}
\let\PRrefmark=\attach
\def\refmark#1{\relax\ifPhysRev\PRrefmark{#1}\else\NPrefmark{#1}\fi}
\def\refend@{\refmark{\number\referencecount}}
\def\refend{\refend@{}\space }
\def\refsend{\refmark{\count255=\referencecount
   \advance\count255 by-\lastrefsbegincount
   \ifcase\count255 \number\referencecount
   \or \number\lastrefsbegincount,\number\referencecount
   \else \number\lastrefsbegincount-\number\referencecount \fi}\space }
\def\refitem#1{\par \hangafter=0 \hangindent=\refindent \Textindent{#1}}
\def\Ref{\rw@trailertrue\REF}
\def\ref{\Ref\?}

\def\REF#1{\r@fstart{#1}%
   \rw@begin{\the\referencecount.}\rw@end}
\def\REFS#1{\r@fstart{#1}%
   \lastrefsbegincount=\referencecount
   \rw@begin{\the\referencecount.}\rw@end}
\def\r@fstart#1{\chardef\rw@write=\referencewrite \let\rw@ending=\refend@
   \ifreferenceopen \else \global\referenceopentrue
   \immediate\openout\referencewrite=referenc.txa
   \toks0={\catcode`\^^M=10}\immediate\write\rw@write{\the\toks0} \fi
   \global\advance\referencecount by 1 \xdef#1{\the\referencecount}}
 {\catcode`\^^M=\active %
 \gdef\rw@begin#1{\immediate\write\rw@write{\noexpand\refitem{#1}}%
   \begingroup \catcode`\^^M=\active \let^^M=\relax}%
 \gdef\rw@end#1{\rw@@end #1^^M\rw@terminate \endgroup%
   \ifrw@trailer\rw@ending\global\rw@trailerfalse\fi }%
 \gdef\rw@@end#1^^M{\toks0={#1}\immediate\write\rw@write{\the\toks0}%
   \futurelet\n@xt\rw@test}%
 \gdef\rw@test{\ifx\n@xt\rw@terminate \let\n@xt=\relax%
       \else \let\n@xt=\rw@@end \fi \n@xt}%
}
\let\rw@ending=\relax
\let\rw@terminate=\relax
\let\splitout=\relax
\def\Closeout#1{\toks0={\catcode`\^^M=5}\immediate\write#1{\the\toks0}%
   \immediate\closeout#1}
%
% % % % % % % % % % % % % % % % % % % % % % % % % % % % % % % % % % % %
%%  Next, figure captions and table captions.
%
\newcount\figurecount     \figurecount=0
\newcount\tablecount      \tablecount=0
\newif\iffigureopen       \newwrite\figurewrite
\newif\iftableopen        \newwrite\tablewrite
\def\FIG#1{\f@gstart{#1}%
   \rw@begin{\the\figurecount)}\rw@end}

\def\Fig{\rw@trailertrue\def\rw@ending{Fig.~\?}\FIG\?}
\def\fig{\rw@trailertrue\def\rw@ending{fig.~\?}\FIG\?}
\def\TABLE#1{\T@Bstart{#1}%
   \rw@begin{\the\tableecount:}\rw@end}
\def\Table{\rw@trailertrue\def\rw@ending{Table~\?}\TABLE\?}
\def\f@gstart#1{\chardef\rw@write=\figurewrite
   \iffigureopen \else \global\figureopentrue
   \immediate\openout\figurewrite=figures.txa
   \toks0={\catcode`\^^M=10}\immediate\write\rw@write{\the\toks0} \fi
   \global\advance\figurecount by 1 \xdef#1{\the\figurecount}}
\def\T@Bstart#1{\chardef\rw@write=\tablewrite
   \iftableopen \else \global\tableopentrue
   \immediate\openout\tablewrite=tables.txa
   \toks0={\catcode`\^^M=10}\immediate\write\rw@write{\the\toks0} \fi
   \global\advance\tablecount by 1 \xdef#1{\the\tablecount}}
%

%
% % % % % % % % % % % % % % % % % % % % % % % % % % % % % % % % % % % %
%%  Finally, inserted figures.
%
%\newread\figureread                                     %% That's
%\def\g@tfigure#1#2 {\openin\figureread #2.fig           %% an example
%   \ifeof\figureread \errhelp=\disabledfigures          %% of \g@tfigure
%     \errmessage{No such file: #2.fig}\let#1=\relax \else
%    \read\figureread to\y@p \read\figureread to\y@p     %%
%    \read\figureread to\x@p \read\figureread to\y@m     %% See LOCPHYX.TEX
%    \read\figureread to\x@m \closein\figureread         %% file for the
%    \xdef#1{\hbox{\kern-\x@m truein \vbox{\kern-\y@m truein
%      \hbox to \x@p truein{\vbox to \y@p truein{        %% actual definition.
%        \special{pos,inc=#2.fig}\vss }\hss }}}}\fi }    %%
%
\def\getfigure#1{\global\advance\figurecount by 1
   \xdef#1{\the\figurecount}\count255=\escapechar \escapechar=-1
   \edef\n@xt{\noexpand\g@tfigure\csname\string#1Body\endcsname}%
   \escapechar=\count255 \n@xt }
\def\g@tfigure#1#2 {\errhelp=\disabledfigures \let#1=\relax
   \errmessage{\string\getfigure\space disabled}}
\newhelp\disabledfigures{ Empty figure of zero size assumed.}
\def\figinsert#1{\midinsert\Tenpoint\medskip
   \count255=\escapechar \escapechar=-1
   \edef\n@xt{\csname\string#1Body\endcsname}
   \escapechar=\count255 \centerline{\n@xt}
   \bigskip\narrower\narrower
   \noindent{\it Figure}~#1.\quad }
%
%%%%%%%%%%%%%%%%%%%%%%%%%%%%%%%%%%%%%%%%%%%%%%%%%%%%%%%%%%%%%%%%%%%%%%%%
%
%   Here come macros for memos & letters.
%
\def\masterreset{\global\pagenumber=1 \global\chapternumber=0
   \global\equanumber=0 \global\sectionnumber=0
   \global\referencecount=0 \global\figurecount=0 \global\tablecount=0 }
\def\FRONTPAGE{\ifvoid255\else\vfill\penalty-20000\fi
      \masterreset\global\frontpagetrue
      \global\lastf@@t=0 \global\footsymbolcount=0}

\def\paperstyle{\letterstylefalse\normalspace\papersize}
\def\letterstyle{\letterstyletrue\singlespace\lettersize}
%  old hoffset = 2 truepc         November 1988
\def\papersize{\hsize=35 truepc\vsize=50 truepc\hoffset=-2.51688 truepc
               \skip\footins=\bigskipamount}
%  old hoffset = 1.3 truein  old voffset = 1.25 truein  November 1988
\def\lettersize{\hsize=5.5 truein\vsize=8.25 truein\hoffset=.4875 truein
	\voffset=.3125 truein
   \skip\footins=\smallskipamount \multiply\skip\footins by 3 }
\paperstyle   %  This is the default
%
% % % % % % % % % % % % % % % % % % % % % % % % % % % % % % % % % % % %
%
\def\MEMO{\letterstyle \letterinfo={\hfil } \let\rule=\memorule
	\FRONTPAGE \memohead }
\let\memohead=\relax

\def\memit@m#1{\smallskip \hangafter=0 \hangindent=1in
      \Textindent{\caps #1}}
\def\subject{\memit@m{Subject:}}
\def\topic{\memit@m{Topic:}}
\def\from{\memit@m{From:}}
\def\to{\relax \ifmmode \rightarrow \else \memit@m{To:}\fi }
\def\memorule{\medskip\hrule height 1pt\bigskip}
\newwrite\labelswrite
\newtoks\rw@toks

\def\addressee#1{\null\vskip .5truein\line{
\hskip 0.5\hsize minus 0.5\hsize\the\date\hfil}\bigskip
   \ialign to\hsize{\strut ##\hfil\tabskip 0pt plus \hsize \cr #1\crcr}
   \writelabel{#1}\medskip\par\noindent}
\def\rwl@begin#1\cr{\rw@toks={#1\crcr}\relax
   \immediate\write\labelswrite{\the\rw@toks}\futurelet\n@xt\rwl@next}
\def\rwl@next{\ifx\n@xt\rwl@end \let\n@xt=\relax
      \else \let\n@xt=\rwl@begin \fi \n@xt}
\let\rwl@end=\relax
\def\writelabel#1{\immediate\write\labelswrite{\noexpand\labelbegin}
     \rwl@begin #1\cr\rwl@end
     \immediate\write\labelswrite{\noexpand\labelend}}
\newbox\FromLabelBox

\let\labelend=\relax
\def\labelbegin#1\labelend{\setbox0=\vbox{\ialign{##\hfil\cr #1\crcr}}
     \MakeALabel }
\newtoks\FromAddress
\FromAddress={}
\def\MakeFromBox#1{\global\setbox\FromLabelBox=\vbox{\Tenpoint
     \ialign{##\hfil\cr #1\the\FromAddress\crcr}}}
\newdimen\labelwidth		\labelwidth=6in
\def\MakeALabel{\vskip 1pt \hbox{\vrule \vbox{
	\hsize=\labelwidth \hrule\bigskip
	\leftline{\hskip 1\parindent \copy\FromLabelBox}\bigskip
	\centerline{\hfil \box0 } \bigskip \hrule
	}\vrule } \vskip 1pt plus 1fil }
\newskip\signatureskip       \signatureskip=30pt
\def\signed#1{\par \penalty 9000 \medskip \dt@pfalse
  \everycr={\noalign{\ifdt@p\vskip\signatureskip\global\dt@pfalse\fi}}
  \setbox0=\vbox{\singlespace \ialign{\strut ##\hfil\crcr
   \noalign{\global\dt@ptrue}#1\crcr}}
  \line{\hskip 0.5\hsize minus 0.5\hsize \box0\hfil} \medskip }
\newbox\letterb@x
\def\lettertext{\par\unvcopy\letterb@x\par}
\def\multiletter{\setbox\letterb@x=\vbox\bgroup
      \everypar{\vrule height 1\baselineskip depth 0pt width 0pt }
      \singlespace \topskip=\baselineskip }
\def\letterend{\par\egroup}
%
%%%%%%%%%%%%%%%%%%%%%%%%%%%%%%%%%%%%%%%%%%%%%%%%%%%%%%%%%%%%%%%%%%%%%%%
%
%   Here come macros for title pages.
%
\newskip\frontpageskip
\newtoks\Pubnum
\newtoks\pubtype
\newif\ifp@bblock  \p@bblocktrue
\def\PH@SR@V{\doubl@true \baselineskip=24.1pt plus 0.2pt minus 0.1pt
             \parskip= 3pt plus 2pt minus 1pt }
\def\PHYSREV{\paperstyle\PhysRevtrue\PH@SR@V}
\def\titlepage{\FRONTPAGE\paperstyle\ifPhysRev\PH@SR@V\fi
   \ifp@bblock\p@bblock \else\hrule height\z@ \relax \fi }
\def\nopubblock{\p@bblockfalse}

\frontpageskip=12pt plus .5fil minus 2pt
\pubtype={\tensl Preliminary Version}
\Pubnum={}
\def\p@bblock{\begingroup \tabskip=\hsize minus \hsize
   \baselineskip=1.5\ht\strutbox \topspace-2\baselineskip
   \halign to\hsize{\strut ##\hfil\tabskip=0pt\crcr
       \the\Pubnum\crcr\the\date\crcr\the\pubtype\crcr}\endgroup}
\def\title#1{\vskip\frontpageskip \titlestyle{#1} \vskip\headskip }
\def\author#1{\vskip\frontpageskip\titlestyle{\twelvecp #1}\nobreak}

\def\address#1{\par\kern 5pt\titlestyle{\twelvepoint\it #1}}
\def\andaddress{\par\kern 5pt \centerline{\sl and} \address}

\def\abstract{\par\dimen@=\prevdepth \hrule height\z@ \prevdepth=\dimen@
   \vskip\frontpageskip\centerline{\fourteenrm ABSTRACT}\vskip\headskip }

%
%
%%%%%%%%%%%%%%%%%%%%%%%%%%%%%%%%%%%%%%%%%%%%%%%%%%%%%%%%%%%%%%%%%%%%%%%%
%   Miscellaneous macros
%

\def\\{\relax \ifmmode \backslash \else {\tt\char`\\}\fi }
\def\sequentialequations{\relax\if\equanumber<0\else\global\equanumber=-1\fi}

\def\journal#1&#2(#3){\unskip, \sl #1\unskip~\bf\ignorespaces #2\rm (19#3),}

\def\topspace{\hrule height 0pt depth 0pt \vskip}

\def\Buildrel#1\under#2{\mathrel{\mathop{#2}\limits_{#1}}}
\def\becomes#1{\mathchoice{\becomes@\scriptstyle{#1}}{\becomes@\scriptstyle
   {#1}}{\becomes@\scriptscriptstyle{#1}}{\becomes@\scriptscriptstyle{#1}}}
\def\becomes@#1#2{\mathrel{\setbox0=\hbox{$\m@th #1{\,#2\,}$}%
	\mathop{\hbox to \wd0 {\rightarrowfill}}\limits_{#2}}}

\let\int=\intop         \let\oint=\ointop
\def\lsim{\mathrel{\mathpalette\@versim<}}
\def\gsim{\mathrel{\mathpalette\@versim>}}
\def\@versim#1#2{\vcenter{\offinterlineskip
	\ialign{$\m@th#1\hfil##\hfil$\crcr#2\crcr\sim\crcr } }}
\def\big#1{{\hbox{$\left#1\vbox to 0.85\b@gheight{}\right.\n@space$}}}
\def\Big#1{{\hbox{$\left#1\vbox to 1.15\b@gheight{}\right.\n@space$}}}
\def\bigg#1{{\hbox{$\left#1\vbox to 1.45\b@gheight{}\right.\n@space$}}}
\def\Bigg#1{{\hbox{$\left#1\vbox to 1.75\b@gheight{}\right.\n@space$}}}
%
% % % % % % % % % % % % % % % % % % % % % % % % % % % % % % % % % % % %
%
%   Finally, some bug fixings.
%
\let\sec@nt=\sec
\def\sec{\relax\ifmmode\let\n@xt=\sec@nt\else\let\n@xt\section\fi\n@xt}
\def\obsolete#1{\message{Macro \string #1 is obsolete.}}
\def\firstsec#1{\obsolete\firstsec \section{#1}}
\def\firstsubsec#1{\obsolete\firstsubsec \subsection{#1}}
\def\thispage#1{\obsolete\thispage \global\pagenumber=#1\frontpagefalse}
\def\thischapter#1{\obsolete\thischapter \global\chapternumber=#1}
\def\REFSCON{\obsolete\REFSCON\REF}
\def\splitout{\obsolete\splitout\relax}
\def\prop{\obsolete\prop \propto }
\def\nextequation#1{\obsolete\nextequation \global\equanumber=#1
   \ifnum\the\equanumber>0 \global\advance\equanumber by 1 \fi}
\def\BOXITEM{\afterassigment\B@XITEM\setbox0=}
\def\B@XITEM{\par\hangindent\wd0 \noindent\box0 }
\def\phyzzx{PHY\setbox0=\hbox{Z}\copy0 \kern-0.5\wd0 \box0 X}
%
%%%%%%%%%%%%%%%%%%%%%%%%%%%%%%%%%%%%%%%%%%%%%%%%%%%%%%%%%%%%%%%%%%%%%%%%
%   That's about it
%
\everyjob{\xdef\today{\monthname\ \number\day, \number\year}}
        
%

%\setup

% Some macros for HEP preprints
% old hoffset=1.0 truein  November 1988
\hoffset=0.2truein
% old voffset=1.0 truein  November 1988
\voffset=0.1truein
\hsize=6truein
\def\TITLEPAGE{\frontpagetrue}
\def\CALT#1{\hbox to\hsize{\tenpoint \baselineskip=12pt
	\hfil\vtop{\hbox{\strut CALT-68-#1}
	\hbox{\strut DOE RESEARCH AND}
	\hbox{\strut DEVELOPMENT REPORT}
	\hbox{\strut hepth@xxx/9212075}}}}

\def\CALTECH{\smallskip
	\address{California Institute of Technology, Pasadena, CA 91125}}
\def\TITLE#1{\centerline{\fourteenpoint #1}}
\def\AUTHOR#1{\vskip .5in \centerline{#1}}

\def\ABSTRACT#1{\vskip .5in \vfil \centerline{\twelvepoint \bf Abstract}
	#1 \vfil}
\def\ENDTITLEPAGE{\vfil\eject\pageno=1}

\def\sqr#1#2{{\vcenter{\hrule height.#2pt
      \hbox{\vrule width.#2pt height#1pt \kern#1pt
        \vrule width.#2pt}
      \hrule height.#2pt}}}
\def\square{\mathchoice\sqr34\sqr34\sqr{2.1}3\sqr{1.5}3}

\def\section#1#2{
\noindent\hbox{\hbox{\bf #1}\hskip 10pt\vtop{\hsize=5in
\baselineskip=12pt \noindent \bf #2 \hfil}\hfil}
\medskip}

\def\underwig#1{	% produce a tilde below the argument
	\setbox0=\hbox{\rm \strut}
	\hbox to 0pt{$#1$\hss} \lower \ht0 \hbox{\rm \char'176}}

\def\bunderwig#1{	% produce a tilde below the argument
	\setbox0=\hbox{\rm \strut}
	\hbox to 1.5pt{$#1$\hss} \lower 12.8pt
	 \hbox{\seventeenrm \char'176}\hbox to 2pt{\hfil}}

\def\MEMO#1#2#3#4#5{
\frontpagetrue
\centerline{\tencp INTEROFFICE MEMORANDUM}
\smallskip
\centerline{\bf CALIFORNIA INSTITUTE OF TECHNOLOGY}
\bigskip
\vtop{\tenpoint \hbox to\hsize{\strut \hbox to .75in{\caps to:\hfil}
\hbox to3in{#1\hfil}
\hbox to .75in{\caps date:\hfil}\quad \the\date\hfil}
\hbox to\hsize{\strut \hbox to.75in{\caps from:\hfil}\hbox to 2in{#2\hfil}
\hbox{{\caps extension:}\quad#3\qquad{\caps mail code:\quad}#4}\hfil}
\hbox{\hbox to.75in{\caps subject:\hfil}\vtop{\parindent=0pt
\hsize=3.5in #5\hfil}}
\hbox{\strut\hfil}}}

\TITLEPAGE
\CALT{1817}
\vskip 1in
\TITLE{Exactly Marginal Operators and}
\vskip 2mm
\TITLE{Running Coupling Constants in 2D Gravity}
\footnote{}{Work supported in part by the U.S. Department of Energy
under Contract No. DE-AC 0381-ER40050.\vskip 0mm}
\vskip .5cm
\AUTHOR{Christof Schmidhuber\footnote*{christof@theory3.caltech.edu}}
\CALTECH
\vskip .5cm
\ABSTRACT{
The Liouville action for two--dimensional quantum
gravity coupled to interacting
matter contains terms that have not been considered
previously.
They are crucial for understanding
the renormalization group flow and
can be observed
in recent matrix model results for the phase diagram
of the Sine--Gordon model coupled to gravity.
These terms ensure, order by order in the coupling constant,
that the dressed interaction is exactly marginal.
They are discussed up to second order.
\vfill
2/93
}
\ENDTITLEPAGE

%References

\Ref\d{F. David, Mod. Phys. Lett. A3, 1651 (1988)}
\Ref\dk{J. Distler and H. Kawai, Nucl. Phys. B 321, 509 (1988)}
\Ref\polch{J. Polchinski, Nucl. Phys. B 346, 253 (1990)}
\Ref\suss{A. Cooper, L. Susskind and L. Thorlacius,
Nucl. Phys. B363, 132 (1991)}
\Ref\poly{A. M. Polyakov, Mod. Phys. Lett. A6, 3273 (1991)}
\Ref\moo{G. Moore, Yale preprint YCTP-P1-92, hepth@xxx/9203061}
\Ref\car{e.g., J. H. Cardy, Les Houches session XLIX, 1988}
\Ref\seg{G. Segal, Comm. Math. Phys. 80, 301 (1981)}
\Ref\wtt{E. Witten, Nucl. Phys. B 373, 187 (1991)}
\Ref\pkl{I. Klebanov and A. M. Polyakov, Mod. Phys. Lett. A6, 3273 (1991)}
\Ref\sei{N. Seiberg, Prog. Theor. Phys. S 102, 319 (1990)}
\Ref\pch{J. Polchinski, Nucl. Phys. 362, 125 (1991)}
\Ref\kut{D. Kutasov, Princeton preprint PUPT-1334 (1992),
hepth@xxx/9207064}
\Ref\pkov{A. M. Polyakov, Lectures in Les Houches, July 1992}
\Ref\gkl{D. Gross and I. Klebanov, Nucl. Phys. B 344, 475 (1990)}
\Ref\mss{G. Moore, N. Seiberg and M. Staudacher,
Nucl. Phys. B 362, 665 (1991)}
\Ref\gou{M. Goulian and M. Li, Phys. Rev. Lett. 66, 2051 (1991)}
\Ref\kle{I. Klebanov, Lectures at ICTP Spring School,
Trieste (1991)}
\Ref\jhs{J. H. Schwarz,
Phys. Lett. B272, 239 (1991)}
\Ref\fts{E.S. Fradkin and A.A. Tseytlin, Nucl. Phys. B 201, 469 (1982)}
\Ref\schm{C. Schmidhuber, Nucl. Phys. B, to appear;
Caltech preprint CALT- 68-1745 (revised 5/92), hepth@xxx/9112005}
\Ref\hsu{E. Hsu and D. Kutasov, Princeton preprint PUPT-1357,
hepth@xxx/9212023 (12/92)}
\Ref\gsw{E.g., M.B. Green, J.H. Schwarz and E. Witten, ``Superstring
Theory", Cambridge University Press 1987}
\Ref\tsey{A.A. Tseytlin, Phys. Lett. B264, 311 (1991)}
\Ref\ban{T. Banks, Nucl. Phys. B 361, 166 (1991)}

{}

\def\bx{{\hbox{ $\sqcup$}\llap{\hbox{$\sqcap$}}}}

\def\sqhat{{\sqrt{\hat g}}}
\def\sqphi{{\sqrt{\hat g}} e^\phi}

{\leftline{\bf 1. Introduction}}

In the Liouville approach,${}^{[\d,\dk]}$
two dimensional quantum gravity coupled
to $c\le 1$ matter `$x$' is formulated in terms of fields propagating
on a fictitious background metric $\hat g_{\alpha\beta}$.
The action is the appropriate conformally invariant free action
plus interaction terms which are usually assumed to be of the form
$${\cal L}_{int}= \hbox{cosmological constant}
+\tau^i\int\Phi_i(x)\ e^{\alpha_i\phi},\eqno(1.1)$$
where $\Phi_i$ are primary fields of the matter theory,
the $\tau^i$ are small coupling constants, $\phi$ is the Liouville mode
and the $\alpha_i$ are adjusted to
make the dimensions of the operators equal to two.

However, (1.1) can not be the complete interaction, for at least two reasons:

1. The operators in (1.1) are not exactly marginal.
\footnote*{An operator is marginal if its dimension is two, and exactly
marginal if its beta function is zero to all orders.}
They should be,
because as a consequence of
general covariance the Liouville theory must be background
independent.${}^{[\d,\dk]}$
Therefore the beta functions
of the theory must be zero to all orders in the couplings.
Adjusting $\alpha_i$ in (1.1) makes them zero to first order,
but whenever there are nontrivial OPE's,
the beta functions have quadratic pieces.
\footnote{\star}{See section 2 for the issue of
renormalization schemes and field redefinitions}

2. The renormalization group flow
would be quite trivial with (1.1). As mentioned, there
should be no flow with respect to the fictitious background scale $\sqhat$.
But, as explained in section 3,
a constant shift of $\phi$ should be interpreted as a rescaling
of the physical cutoff,${}^{[\polch,\suss,\poly]}$
and should in particular result in a mixing (flow) between
different operators. This does not happen in (1.1).

As shown in section 2, the first problem
can be solved by adding a term
$$\propto\ \ \
-c^k_{ij}\tau^i\tau^j\int \Phi_k(x)\ \phi\ e^{\alpha_k\phi}\eqno(1.2)$$
to the interaction (1.1) where $c^k_{ij}$ are the operator
product coefficients.
This, in fact, also resolves the second problem: the modified
interaction displays the expected operator mixing
under shifts of $\phi$ by a constant.
Requiring that there be {\it no} flow w.r.t.
the background scale
$\sqrt{\hat g}$ determines the
flow w.r.t. the physical scale ${\sqrt{\hat g}}e^{\alpha\phi}$.
For the case of the Sine--Gordon model coupled to gravity it will be seen
that this
flow qualitatively agrees with recent matrix model results
by Moore.${}^{[\moo]}$

(1.2) should be viewed as a second order
correction to the gravitational dressing of the
$\Phi_i(x)$.
It is conjectured that further modifications of (1.1)+(1.2)
can be made order by order in the $\tau^i$, leading to an infinite
dimensional space of exactly
marginal perturbations.

The paper is organized as follows:

In section 2, the second order corrections (1.2)
are discussed. First, it is shown in subsection 2.1 that
the interaction (1.1) plus (1.2) is marginal up to second order.
That the correction (1.2) is essentially unique is argued
in appendix A by thinking of the marginality conditions
as equations of motion of string theory.
The $c=1$ model coupled to gravity is
discussed as an example, the interaction terms being
the Sine--Gordon model near
the Kosterlitz-Thouless momentum $p={\sqrt2}$
and near $p={1\over2}\sqrt2$ in subsection 2.2,
and the ``discrete operators'' in subsection 2.3.
The effects of including the cosmological constant
are studied in appendix B.
The conclusions of appendices A and B are summarized in
subsection 2.4.

In section 3, running coupling constants are discussed.
They are defined in subsection 3.1 so that they
absorb a constant
shift of $\phi$. In subsections 3.2 and 3.3
this is applied to the Sine-Gordon model and the resulting phase
boundaries are compared with those found with the
nonperturbative matrix model
techniques.${}^{[\moo]}$
It is seen that
the presence of the terms (1.2) is crucial even for qualitative
agreement of the matrix model-- and the Liouville
approaches. A more detailed comparison
of both is left for later.
The one--loop beta functions for the discrete $c=1$ operators
are also obtained.

In section 4,
possible extensions of this work are pointed out, as well as
implications for black hole hair and correlation functions.
In particular, it is argued that the relation between
correlation functions in the matrix model and in the Liouville
approach is more complicated than often assumed.

\vfill\eject

\def\h{h_{\mu\nu}}

\def\t{\cos px\ e^{(p-{\sqrt2})\phi}}
\def\tt{\tau}

\def\stwo{{\sqrt2}}

{\leftline{\bf 2. Exactly Marginal Operators}}

$\underline{\hbox{2.1. The Terms of Order $\tau^2$}}$

In DDK's
\footnote{\bullet}{David, Distler and Kawaii}
approach,
a conformal field theory with central charge
$c$ and Lagrangian ${\cal L}_m(x)$
coupled to 2D gravity is described by the action${}^{[\d,\dk]}$
$$S_0={1\over{8\pi}}\int\sqhat\{ {\cal L}_m(x)+\partial\phi^2+Q
\hat R\phi+\hbox{cosmological constant}+\hbox{ghosts}\}\eqno(2.1)$$
with $Q={\sqrt{(25-c)/3}}$ and conformal factor $\phi$. The
cosmological constant will be neglected first, but included later.
(see subsection 2.4.)

When ${\cal L}_m(x)$ is
perturbed by operators $t^i\Phi_i(x)$, these operators get
``dressed'' upon coupling to gravity.
As mentioned in the introduction, the dressed interaction
must be an {\it exactly}
marginal operator, not only an operator of dimension two.
Exact marginality
is needed because in DDK's approach the
background metric $\hat g$ is an arbitrary gauge choice that
nothing should depend on.
In particular, coupling constants
should not run with respect to $\hat g$: all beta functions must be
zero to all orders.

So far, this condition has been exploited only
to first order${}^{[\d,\dk]}$.
Here it will be exploited up to second order.
Generally,${}^{[\car]}$
the beta functions for a perturbed conformally
invariant theory $S_0$,
$$S=S_0+\tau^i\int d^2r\ V_i,
\footnote*{\hbox{Here and below we omit powers of a length scale
$a$, needed to make the $\tau^i$ dimensionless.}}$$
$$\hbox{are}\ \ \ \ \ \beta^i=(\Delta^i_j-2\delta^i_j)\ \tau^j+\pi c^i_{jk}
\tau^j\tau^k+o(\tau^3),\eqno(2.2)$$
if the $V_i$ are primary fields of dimension $\Delta_i$ close to two.
$\Delta^i_j$ is the dimension matrix computed with $S_0$.
If the operators $V_k$ on the RHS of the operator algebra
\footnote\square{keeping only the radial dependence on the RHS;
the rest drops out after
integrating over $\vec r$.}
$$V_i(r)V_j(0)\sim \sum_k \vert r\vert^{-\Delta_i-\Delta_j+\Delta_k}
c^k_{ij} V_k(0)$$
also have dimension
close to two , the coefficients $c^k_{ij}$ are universal constants,
independent of the renormalization scheme used to compute them.
Operators of other dimensions also appear on the RHS.
For them, the
$c^i_{jk}$ are scheme--dependent, that is, not invariant under
coupling constant redefinitions. Let us ignore them here.
\footnote{\star}{
Presumably the scheme can be chosen so that they vanish.}

We now show that $\beta^i=0+o(t^3)$ for the perturbation (1.1) plus
(1.2):
\footnote\circ{The question of the uniqueness
of (1.2) is deferred to subsection 4.4.}
$$\delta S\ =\ \tau^i\int V_i(x,\phi)\ \equiv \ \tau^i\int
 \hat V_i(x,\phi)
-\pi\ c^k_{ij}\tau^i\tau^j\int X_k(x,\phi),\eqno(2.3)$$
$$V_i=\hat V_i-\pi c^k_{ij}\tau^jX_k,\ \ \ \ \hat V_i \equiv
\Phi_i(x)\ e^{\alpha_i\phi},\ \ \ \
X_k\equiv -{1\over{Q+2\alpha_k}}\ \Phi_k(x)\ \phi\ e^{\alpha_k\phi}.
\eqno(2.4)$$
$\alpha_i$ is adjusted to make the dimension of $\hat V_i$
exactly two. Without the $o(\tau)$--corrections in $V_i$, we would thus
have
$\Delta^i_j=2\delta^i_j$ and
$\beta=0+o(\tau^2)$ from (2.2). With them,
$$\Delta^i_j=2\delta^i_j-\pi c^i_{kj}\tau^k+o(\tau^2),\eqno(2.5)$$
hence $\beta=0+o(\tau^3)$ in (2.2). (2.5) can be derived by
writing
$$X_k =\ -{1\over{Q+2\alpha_k}}\ \Phi_k(x)\ {\partial\over{\partial\alpha_k}}
\ e^{\alpha_k\phi},$$
defining the generator $L_0+\bar L_0$ of global scale transformations
and differentiating with respect to $\alpha_k$ the dimension formula
$$\eqalign{
&(L_0+\bar L_0) e^{\alpha_k\phi} = -\alpha_k(\alpha_k+Q)\ e^{\alpha_k\phi},\cr
\Rightarrow\ \ \ &(L_0+\bar L_0) X_k= 2X_k+ \hat V_k\cr
\Rightarrow\ \ \ &(L_0+\bar L_0) V_k= 2V_k-\pi c^i_{jk}\tau^jV_i+o(\tau^2).}$$

As a simple check of all this, one can consider rescaling
$\psi\rightarrow (1+\lambda)\psi$ in
$$S_{\hbox{ toy model}}\ =
\ {1\over{8\pi}}\int\partial \psi^2+\gamma
\cos\stwo \psi\ \ \ \hbox{with}\ \ \
\lambda\ll \gamma\ (\cdot a^2).$$
This should keep
the interaction marginal at $o(\gamma\lambda)$ and is equivalent to
adding the terms $2\lambda\ \partial \psi^2
-\stwo\lambda \gamma\ \psi\sin\stwo \psi$ to
the Lagrangian. Using the above method, one can check
that the second term indeed arises as the correction to the first term.
\footnote\odot{Here, $8\pi \hat V_1= \cos\ \stwo\psi,\ 8\pi\hat V_2
= 2\partial \psi^2$,  $c^1_{12}=c^1_{21}=-2/\pi$ and
$8\pi X_1=-1/(2\stwo)\ \psi\sin\ \stwo\psi$.}

$\underline{\hbox{2.2. The Sine--Gordon model}}$

As an example, consider an uncompactified
scalar field $x$ coupled to gravity. Then $c=1$ and $Q=2\stwo$. First,
we perturb this model
by the Sine--Gordon interaction,
$$S={1\over{8\pi}}\int\sqhat\{ \partial x^2+\partial\phi^2+2{\sqrt2}
\hat R\phi+\hbox{ghosts}\}+
m\int\t,\eqno(2.6)$$
and determine the $o(m^2)$ corrections (2.3).
To find the coefficients $c^k_{ij}$, consider
the operator product expansion (OPE), using the propagator $-\log r^2$:
$$\eqalign{
\cos px\ &e^{(p-{\sqrt2})\phi}(r)\ \
\cos px\ e^{(p-{\sqrt2})\phi}(0)\cr
&\sim \vert r \vert
^{-2-4(p-{1\over2}{\sqrt2})^2}\ e^{2(p-{\sqrt2})\phi}
\ \{{1\over2}-\vert r\vert^2
 {{p^2}\over8}\partial x^2
+...\}\cr &+\vert r\vert
^{-2+(4{\sqrt2}p-2)}\ \cos 2px\
e^{2(p-{\sqrt2})\phi}
\ \{{1\over2}-\vert r\vert^2
\ {{p^2}\over8}\partial x^2+...\}
}\eqno(2.7)$$
As mentioned, we must look for nearly quadratic singularities,
so that the $c^i_{jk}$ are universal constants.
The second line in (2.7)
has quadratic singularities at the ``discrete momenta''
$p\in\{...,0,{1\over2}\stwo,\stwo,...\}$
\footnote*{corresponding to the discrete tachyons $\Phi_{j,\pm j}$ of the
next subsection}
and the third line at
$p\in\{...,0,{1\over4}\stwo.\}$
\footnote{\star}{However, for $p<{1\over2}\stwo$ the operators
on the RHS do not exist.
As a consequence, $\cos 2px-$terms will not be induced and
nothing happens at those momenta, as argued in
appendix B.}
Let us study the neighborhoods of $p={1\over2}\stwo$
and $p=\stwo$. There the induced operators are:
$$\eqalign{\hbox{at}\ p={1\over2}\stwo+\delta:\ \ \
& \hat V_1=
e^{(2\delta-\stwo)\phi}\ \ \ \hbox{with}\ c^1_{mm}={1\over2}\cr
\hbox{at}\ p=\stwo+\epsilon:\ \ \
& \hat V_2=\partial x^2\ e^{2\epsilon\phi}\ \ \ \hbox{with}\ c^2_{mm}
=-{{p^2}\over8}}$$
{}From (2.3) and (2.4), the leading order corrections to (2.6)
are obtained:
$$\eqalign{
&\hbox{near } p={1\over2}\stwo:\ \ \delta S=
 {{m^2\pi}\over{8\delta}}\
\int\phi\ e^{-\stwo\phi} \cr
&\hbox{near } p=\stwo:\ \ \delta S=
-{m^2\pi\over{8\stwo}}\int\phi\ \partial x^2
}\eqno(2.8)$$
This will be further discussed in section 3. Note that from
the string theory
point of view, (2.8) describes the backreaction of the tachyon
onto itself and the graviton.

$\underline{\hbox{2.3. The Discrete $c=1$--Operators}}$

As a second example, consider perturbing the $c=1$ model with
the nonrenormalizable so--called
(chiral) discrete primaries $\Phi_{jm}(x)$,${}^{[\seg]}$
$$\Phi_{jm}=\ f_{jm}[\partial x,\partial^2 x,...]
e^{im{\sqrt 2}x}\ \equiv (\oint e^{-i{\sqrt 2}x})^m\ e^{ij{\sqrt 2}x}
\eqno(2.9)$$
with dimension $j^2$ and
$SU(2)$ indices $j,m$, the $SU(2)$ algebra being generated by
$$H^{\pm}\sim\ \oint\ e^{\pm i{\sqrt 2}x}\ =\oint\ \Phi_{1,\pm1},
\ \ \ H^3\sim
\oint\ i\stwo\partial x=\oint\ \Phi_{1,0}.$$
If an interaction
$t^{jm}\Phi_{jm}[x]\bar\Phi_{jm}[\bar x]$
is added to the matter Lagrangian, the dressed interaction is given to
first order in the coupling constants by
$${\cal L}_{int}=
\tau^{jm}\ \hat V_{jm},\ \ \ \hat V_{jm}\equiv  \Phi_{jm}(x)
\bar\Phi_{jm}(\bar x)\ e^{\alpha_j(\phi+\bar\phi)}$$
with $\alpha_j=(j-1){\sqrt 2}$. The $\Phi_{jm}$ can be rescaled
such that the operator algebra of the
$\hat V_{jm}$ has the
$w_\infty$
structure${}^{[\wtt,\pkl]}$
$$c^{jm}_{kn\ k'n'}=(kn'-k'n)^2\ \delta_{j,k+k'-1}\delta_{m,n+n'}.$$
{}From (2.3) and (2.4) one obtains the second order interaction term
$$\delta {\cal L}=
\sum_{j,m}\ \Phi_{jm}\bar\Phi_{jm}\ \phi\ e^{\alpha_j\phi}\ \
\times\ {\pi\over{2\stwo j}}
\sum_{j'+j''=j+1\atop m'+m''=m}
(j'm''-j''m')^2\ \tau^{j'm'}\tau^{j"m"}.
\eqno(2.10)$$
${\cal L}_{int}+\delta {\cal L}$ is marginal up
to order $\tau^2$.
Again, with different renormalization schemes operators whose dimension
is not two may also appear in $\delta {\cal L}$.

$\underline{\hbox{2.4. Uniqueness and the Cosmological Constant}}$

Next, we must ask whether the modifications (1.2) of the operators
(1.1) are
the unique modifications that achieve marginality
up to order $\tau^2$.
The situation
is greatly clarified by thinking of the marginality conditions
as equations of motion of string
theory, as in ref. [\polch].
One concludes the following (more details are in appendix A):

The marginality conditions are second order differential equations
in $\phi$ and $x$. Their solutions are unique after two boundary
conditions are imposed, namely:
(i): the modifications must
vanish at $\phi=0$, and (ii): the second, more negative
Liouville dressing (as e.g., in (A.5)) does not appear.
Boundary condition (ii) arises because operators with the more
negative Liouville dressing
do not exist.

Boundary condition (i) comes about because
the Liouville mode
$\phi$ lives on a half line:${}^{[\polch]}$
The sum over geometries can be covariantly regularized as a sum over random
lattices. Then no two points can come closer to each other than the
lattice spacing $a$:
$$\hat g_{\mu\nu}\ e^{\alpha\phi}\ d\xi^\mu d\xi^\nu\ge a^2\ \ \
\Rightarrow\ \ \ \phi\le\phi_0\ \ \hbox{with}\ \
e^{\alpha\phi_0}\propto a^2\eqno(2.11)$$
(recall that $\alpha<0$.)
After shifting $\phi$ so that $\phi_0=0$, boundary condition
(i) states that the action $S(\phi=0)$ at the cutoff scale is the bare
action (see e.g., (A.4)).${}^{[\polch]}$

The correction terms found above
obey the boundary conditions (i) and (ii) and are therefore
unique.
Of course, there is always an ambiguity due to field redefinitions,
that is, to choosing different renormalization schemes when
computing the beta function.
There is no problem as long as we stick to one scheme.
\footnote*{
Actually, the scheme used subsection 2.1.
is not the same as the one used for the string equations of motion
in appendix A, but this does not affect the above conclusions.}

Another important question is
how the cosmological constant
modifies our results.
The problem with the cosmological constant operator is that it cannot be made
small in the IR ($\phi\rightarrow -\infty$). It can only be
shifted in the $\phi$--direction.
Thus it cannot be treated perturbatively, rather it should be included
from the start in $S_0$ of (2.1). In its presence
the OPE's used above are modified. Applying
the discussions in refs. [\polch,\sei,\pch],
one tentatively
concludes the following (more details are in appendix B):

1. The effects of the
cosmological constant on gravitiational dressings can be
neglected in the ultraviolet ($\phi\sim0$), but not in the
infrared ($\phi\rightarrow -\infty$).

2. In the Sine--Gordon model coupled to gravity, no unwanted
terms with $\cos 2px$ are induced because the OPE's
are ``softer'' than in free field theory (see (A.4-5)).

This will be
confirmed in section 3 by the agreement with
matrix model results.

\vfill\eject

{\leftline{\bf 3. Running Coupling Constants}}

$\underline{\hbox{3.1. Renormalization Group Transformations}}$

In subsections 3.1 and 3.2, the cosmological term will be assumed to be
$\mu\ e^{-\stwo\phi}$ to simplify the discussion, which can be
generalized to more complicated forms like $T_\mu(\phi)$ in (B.2).

Consider rescaling the cutoff
$a\rightarrow ae^\rho$
in the path integral of 2D gravity,
$$\int_{\phi\ge\phi_0} D\phi\ Dx\ Db\ Dc\ e^{-S(\phi,x,b,c)}$$
{}From (2.11) one sees that this induces a shift
of the bound $\phi_0\rightarrow
\phi_0+\lambda,$ in addition to an
ordinary RG transformation. Here,
$$\lambda=\phi_0(a e^\rho)-\phi_0(a)={2\over\alpha}\rho=-\stwo\rho.
\footnote{\oplus}{\hbox{More generally,
$\rho={1\over2}\log{{T_\mu(\lambda)}\over{T_\mu(0)}}$
with cosmological constant $T_\mu$ as in (B.2)}}\eqno(3.1)$$
In fact, since ordinary RG transformations
are irrelevant (all beta functions are zero),
$only$ the shift of the bound remains.
The constant shift of the
bound is equivalent to a constant shift of the Liouville mode,
$\phi\rightarrow \phi+\lambda$.
Let us absorb this shift in
``running coupling constants'' $\vec\tau(\lambda),\vec\tau_0
\equiv \vec\tau(0)$,
defined by:
$$S[\vec\tau(\lambda),x,\phi+\lambda]=S[\vec\tau_0,x,\phi].\eqno(3.2)$$
After expressing $\lambda$
in terms of $\rho$,
\footnote{\otimes}{This is more complicated with $T_\mu$.
But to find phase boundaries,
$\vec\tau(\lambda)$ will be good enough.}
one obtains the renormalization group
flow $\vec\tau(\rho)$ (for similar conclusions, see
[\suss,\kut].)

As mentioned above,
the action (3.2) corresponds to a classical solution of string theory
with two--dimensional target space ($x,\phi$).
The equations of motion of classical string theory thus play the role
of the Gell--Mann--Low equations in the presence of gravity.${}^{[\suss]}$
They contain
second (and higher) order derivatives of $\phi$, i.e. $\rho$.
It has been suggested that those
are due to the contribution of pinched
spheres in the functional integral over metrics.${}^{[\pkov]}$
\vskip1cm
$\underline{\hbox{3.2. Sine--Gordon Model near $p=\stwo$}}$

We now apply the preceding to the examples worked out in section 2,
starting with the Sine--Gordon model.
In flat space, at $p=\stwo$ the Kosterlitz-Thouless phase transition occurs.
With gravity, at $p=\stwo+\epsilon$ the action
is to order $(m,\epsilon)^2$
(see (2.8); we ignore $o(\mu)$--corrections
of the Sine--Gordon interaction):
$$\eqalign{S={1\over{8\pi}}\int\sqhat&\{ \partial x^2+\partial\phi^2+2{\sqrt2}
\hat R\phi
+\hbox{ghosts}+\mu\ e^{-\stwo\phi}\}\cr
&+m\int\cos(\stwo+\epsilon)x\ e^{\epsilon\phi}
-{\pi\over{8\stwo}}\ m^2\int\phi\ \partial x^2
}\eqno(3.3)$$
To $o(m,\epsilon)^2$, a shift
$\phi\rightarrow\phi+\lambda$ can be absorbed in the
$\lambda$--dependent couplings
$$m(\lambda)=m_0e^{-\epsilon\lambda},\ \ \
\epsilon(\lambda)=\epsilon_0-{\pi^2\over{2}}\lambda m^2,\ \ \
\mu(\lambda)=\mu_0e^{\stwo\lambda}.
$$
In deriving $\epsilon(\lambda)$,
the $\lambda m^2\partial x^2$ term has been absorbed
in a redefinition of $x$ and then in a shift of $\epsilon$. Defining
`prime' as ${d\over{d\lambda}}$, we get
$$\epsilon'=-{\pi^2\over{2}}m^2+..,\ \ \
m'=-\epsilon m+..,\ \ \
\mu'=\stwo\mu+..
$$
Defining `dot' as ${d\over{d\rho}}
=-\stwo{d\over{d\lambda}}$ yields the
beta functions
$$\dot\epsilon={\pi^2\over{\stwo}}m^2,
\ \ \ \dot m={\stwo}\epsilon m,\ \ \ \dot\mu=-2\mu.\eqno(3.4)$$
$\dot\mu$ serves as a check: $\mu$ decays in the UV according
to its dimension two.
The coupling constant flow is qualitatively the same as in flat
space and is given by the Kosterlitz-Thouless diagram
(Figure 1). We see that the $m^2 \phi\ \partial x^2$
correction (1.2) of (1.1) plays a crucial role:
ignoring it would be like forgetting about field renormalization
in the ordinary Sine-Gordon model.

{}From (3.4), the phase boundary for $p>\stwo$ is linear,
$m \propto \epsilon$.
To this order, this agrees with the matrix model result ${}^{[\moo]}$
$$m\ \propto\ \epsilon\ e^{{1\over2}\stwo\epsilon
\log\epsilon}.\eqno(3.5)$$
With the normalization of $m$ and $\epsilon$ as in (3.3), we obtain
the slope $\stwo/\pi$ for the phase boundary. After comparing
the normalizations, this should also
be checked with the matrix model.
It will also be interesting to see if the logarithm in
(3.5)
follows from the
modifications of higher order in $m$, needed to keep the interaction
near $p=\stwo$ marginal beyond $o(m^2)$.

We can now interpret the phase diagram of [\moo]
(figure 2) near $m,\epsilon=0$:
For $\epsilon<0$ (regions II and V of [\moo]),
$m$ grows exponentially towards the IR.
The model thus flows to (infinitely many copies of) the $c=0$,
pure gravity model.${}^{[\kut,\gkl]}$

For $\epsilon>0$ but $m$ greater than a critical value $m_c(\epsilon)$
(region VI of [\moo],) we flow again towards
the $c=0$ model in the IR. For $m<m_c(\epsilon)$ (region III of [\moo])
the flow seems to take us to the free $c=1$ model. However,
the domain of small $\epsilon,m$ is now the IR domain. As noted
in subsection 2.4, the cosmological
constant can not be neglected there and further investigation is needed.

\vskip5mm
$\underline{\hbox{3.3. Sine--Gordon Model near $p={1\over2}\stwo$}}$

At $p={1\over2}\stwo+\delta$, the situation is less clear.
{}From (2.8),
instead of the $\partial x^2$--term a ``1''--term
is induced. That is,
the cosmological constant
is modified by the induced
operator $\phi\ e^{-\stwo\phi}$. The latter becomes comparable with
the background cosmological constant at $\delta\sim{{m^2}\over\mu}$.
Let us tentatively
\footnote\bullet{Here we use $\phi\ e^{-\stwo\phi}$ instead of the
simple form $e^{-\stwo\phi}$ for the cosmological
constant.}
write the action
to leading order as:
$$\eqalign{S={1\over{8\pi}}\int\sqhat&\{
\partial x^2+\partial\phi^2+2{\sqrt2}
\hat R\phi
+\hbox{ghosts}\}\cr
&+m\int\cos({1\over2}\stwo+\delta)x\ e^{(-{1\over2}\stwo+\delta)\phi}
+({\mu\over{8\pi}}+{{m^2}\pi\over{8\delta}})\int \phi\ e^{-\stwo\phi}
}\eqno(3.6)$$
With our normalizations, the effective cosmological constant is now
$\mu+{{m^2}\over{\delta}}\pi^2$. For fixed $\mu$, it blows up
as $\vert\delta\vert\rightarrow 0$.
For $\delta<0$ and
$m\ge{1\over\pi} \sqrt{\vert\mu\delta\vert}$, it is negative.
Indeed, in the matrix model a singularity of the free energy has been found
at
$$\delta<0,\ \ \ \
m\propto{\sqrt{\vert\mu\delta\vert}} e^{{1\over2}\stwo \delta\log\delta}.
\eqno(3.7)$$
Let us therefore identify the region
where $\mu+{{m^2}\over{\delta}}\pi^2$ is negative with
region IV of [\moo].
We leave a further interpretation of the situation
near $p={1\over2}\stwo$ for the future.

$\underline{\hbox{3.4. The Discrete Operators}}$

We can also determine the one--loop beta function
for the ``discrete'' interactions (2.9)
of subsection 2.3. From (2.10),
$${\cal L}+\delta {\cal L}=\sum_{j,m}
\Phi
\bar\Phi_{jm}\ e^{(j-1)\stwo\ \phi}
\{\tau^{jm}+\phi\
 {\pi\over{2\stwo j}}
\sum_{j'+j''=j+1\atop m'+m''=m}
(j'm''-j''m')^2\tau^{j'm'}\tau^{j"m"}\}.$$
Constant shifts $\phi\rightarrow\phi+\lambda$
are absorbed up to $o(\tau^2)$ in:
$$\tau^{jm}(\vec \tau_0,\lambda)=\{\tau^{jm}_0
-\lambda\times {\pi\over{{2\sqrt 2 }j}}
\sum_{j'+j''=j+1\atop m'+m''=m}
(j'm''-j''m')^2
\tau^{j'm'}_0\tau^{j"m"}_0
\} e^{-(j-1)\stwo \hbox{$\lambda$}}
.$$
{}From this we find the one loop beta function (using (3.1)):
$$\dot \tau^{jm}
=2(j-1)\ \tau^{jm}
+{\pi\over{2j}}\sum_{{j'+j''=j+1}\atop{m'+m''=m}}
(j'm''-j''m')^2\ \tau^{j'm'}\tau^{j"m"}
\ +o(\tau^3).
\eqno(3.8)$$
Thus, turning on
the operators $\Phi\bar\Phi_{j'm'}$ with $j'>1$ will in general
induce an infinite set of higher spin operators $\Phi\bar\Phi_{jm}$ at
$o(\tau^2)$, whose couplings
were originally turned off.
This is what one expects from these nonrenormalizable
operators, but it would not happen without
the $o(\tau^2)$ modification $\delta{\cal L}$.

\vfill\eject

{\leftline{\bf 4. Outlook}}

$\underline{\hbox{4.1. Correlation Functions}}$

The modifications (1.2) are important not only for understanding
the renormalization group flow but also
for computing correlation functions in Liouville theory.
They imply the identification (the notation is as in (2.4)):
$$<\exp\{\int t^i \Phi_i\}>_{G}\ \sim\
<\exp\{\int(\tau^i\ \hat V_i+\kappa_l c^l_{ij}\tau^i\tau^j
\phi \hat V_l+..
)\}>_{L}.\eqno(5.1)$$
where $<...>_{G}$ and $<...>_{L}$ denote correlation functions
computed in the matrix model (Gravity)
and in Liouville theory, respectively, and
$\kappa_l=\pi/(Q+2\alpha_l)$.
$\tau^i$ is related to the $t^j$ in some nontrivial way.${}^{[\mss]}$

Geometrically, the extra terms on the RHS can be interpreted as arising from
pinched spheres in the sum over surfaces.
(5.1) has consequences for the correspondence of matrix model and
Liouville correlation functions. Expanding both sides and temporarily
identifying $t$ and $\tau$
\footnote\star{
The nontrivial relation between $\tau$ and $t$ noted in [\mss] corresponds
to the appearance of operators $\hat V_l$ (instead of $\phi \hat V_l$) on the
RHS of (5.1). They are also present, but let us here focus on
the new type of operators $\phi \hat V_l$.}
yields, e.g, for the two-point function:
$$<\int\Phi_i\int\Phi_j>_{G}\ =\int d^2z\int d^2w<\hat V_i(z)\hat V_j(w)>_{L}
+\ 2\kappa_l
c^l_{ij}\int d^2w<\phi\ \hat V_l(w)>_{L}
\eqno(5.2)$$
In fact, the last term is necessary for background invariance: Inserting
a covariant regulator
like $\Theta(\sqphi\vert z-w\vert^2-a^2)$
into the two-point function
induces new background dependence, coming from the
integration region $z\sim w$ ${}^{[\car]}$.
By construction, the one--point functions added in (5.2) are precisely
the ones needed to cancel this dependence.

Analoguously,
additional terms like the ones in (5.2) are
also present in higher point functions. They can be determined by
background invariance. It should be possible to see them in matrix model
computations, e.g., of higher--point functions of tachyons at the
``disctrete'' momenta.
It then needs to be better understood
why we can recover some of the matrix model correlators
with the method of
Goulian and Li
from the Liouville correlators,${}^{[\gou,\kle]}$
without the
extra terms in (5.2).
\vskip 1cm

$\underline{\hbox{4.2. Black Hole Hair}}$

The conjecture that all the discrete operators, in particular the
`static' ones $\Phi_{j,0}$ with zero $x$--momentum can be turned into exactly
marginal ones implies that
each $\Phi_{j,0}$ adds
a new dimension to the space of black hole solutions of
classical 2D string theory, corresponding to higher spin (not
only metric) hair.
It will be very interesting to better understand in how far
this is significant for the issue of information loss
in black holes.${}^{[\jhs]}$

\vskip 1cm
$\underline{\hbox{4.3. Four Dimensions}}$

It would also be interesting to extend this work to four
dimensions. Four--dimensional Euclidean quantum geometry
is, at the least, an interesting statistical mechanical model.
At the ultraviolet fixed point of infinite Weyl coupling,
where the theory is asymptotically free${}^{[\fts]}$
it can be solved
with the methods of two dimensional quantum
gravity in conformal gauge.${}^{[\schm]}$
Perturbing away
from this limit is similar to adding perturbations to
the free $c=1$ theory. One might be able to find a phase diagram for
Euclidean quantum gravity
by generalizing the method suggested in this paper.

\vfill\eject
$\underline{\hbox{4.4. Summary}}$

In the Liouville theory approach to 2D quantum gravity coupled to an
interacting scalar field, new terms
appear in the Lagrangian at higher orders in the coupling constants.
They are required by background invariance and
cannot be eliminated by a field redefinition when the interaction
is given by one of the discrete tachyons or higher--spin operators.

The new terms are crucial for obtaining
the correct phase diagram, as found with
the nonperturbative matrix model techniques in the
case of the Sine--Gordon model. We have partly interpreted
this diagram, but the transition below
${1\over2}$ of the Kosterlitz--Thouless momentum must be clarified more,
the cosmological constant must be treated more rigorously, and
the cubic terms in the beta function (2.2), which are also universal,
should be derived. The new terms have various other implications
and should in particular be important for the correct computation
of higher--point correlation
functions.

\vskip 1in

{\bf{Acknowledgements}}

I would like to thank J. H. Schwarz for
questions and advice, A. M.
Polyakov for a
discussion in Les Houches and K. Li for critical comments.
I also thank D. Kutasov and many others for comments on the first
draft.

\vskip1cm
 {\bf{Note added:}}

Some other aspects of the Sine--Gordon model
coupled to gravity have been
studied recently in ref. [\hsu].

\vfill\eject

$\underline{\hbox{Appendix A:
String Equations of Motion and Boundary Conditions}}$

The question is whether (1.2) are the unique modifications that
make the interaction (1.1) marginal up to order $\tau^2$.
It is useful to think of 2D quantum gravity as classical
string theory.${}^{[\polch]}$
Let us first discuss the example of
the Sine--Gordon model.
The discussion will be restricted to genus zero.

It is well known that, for genus zero, exactly marginal
perturbations of the world sheet action
correspond to classical solutions of string theory.
Some of them can be found by expanding
the dilaton $\Phi$, the graviton $G_{\mu\nu}$
and the tachyon $T$ in the sigma model
$$S={1\over{8\pi}}\int\sqhat\{
G_{\mu\nu}(x,\phi)\partial x^\mu\partial x^\nu
+\hat R\Phi(x,\phi)+T(x,\phi)\}\hskip1cm\hbox{in m:}
\footnote\triangleright{\hbox{The cosmological
constant will be included in the tachyon
in appendix B.}}
\footnote{}{\hbox{
Its presence justifies the expansion in $m$.}}
\eqno(A.1)$$
$$\eqalign{
T(x,\phi)&
=m\t+m^2T^{(1)}(x,\phi)
+..\cr
\Phi(x,\phi)&
=2{\sqrt2}\phi+m^2\Phi^{(1)}(x,\phi)+..\cr
G_{\mu\nu}(x,\phi)&
=\delta_{\mu\nu}+m^2\h(x,\phi)+..
}\eqno(A.2)$$
and by then solving the equations of motion derived from
the low-energy effective action of two-dimensional string
theory,${}^{[\gsw,\tsey]}$
$$\int dx\ d\phi\ {\sqrt G}e^\Phi
\{R+\nabla\Phi^2+8+\nabla T^2-2T^2+{4\over3}T^3+o(m^4)
\}.$$
The corrections to $G,\Phi$ in (A.2) are of order $m^2$ because $T$ appears
in the Hilbert-Einstein equations only in the tachyon stress tensor, which
is quadratic in $T$. The $T^3$--term is ambiguous,${}^{[\tsey,\ban]}$
but this will not be important here.
It is useful to choose a gauge in which the dilaton is linear,
i.e., $\Phi^{(1)}=0$.
\footnote{\triangleleft}{This
is always possible at least at order $m^2$ and $m^3$.}
To $o(m^2)$,
the equations of motion are second order differential equations:
$$\eqalign{&\nabla_\mu\nabla_\nu\Phi-{1\over2}G_{\mu\nu}
({\vec\nabla}\Phi^2+2\bx\Phi-8)=\Theta_{\mu\nu}\cr
&\bx T+{\vec\nabla}\Phi{\vec\nabla}T+2T-2T^2=0
}\eqno(A.3)$$
with tachyon stress tensor $\Theta_{\mu\nu}$.

To specify a
solution, we need two boundary conditions.
Following ref. [\polch], we will adopt boundary conditions
given (i) in the
ultraviolet
by the bare action and (ii) in the infrared
by the requirement
of regularity.
Let us for now assume the simple form
$e^{\alpha\phi}$ for the cosmological constant. `Infrared' means
$\phi\rightarrow -\infty$
since $\alpha=-\stwo<0$.

(i) UV: As pointed out in (2.11), the Liouville coordinate is bounded:
$$\hat g_{\mu\nu}\ e^{\alpha\phi}\ d\xi^\mu d\xi^\nu\ge a^2\
\Rightarrow\ \ \ \phi\le\phi_0\sim {1\over\alpha}\log a^2.$$
This bound on $\phi$ does not modify the Einstein equations.
\footnote{\diamond}{The implicit assumption here is that the term
$\log\sqrt{ \hat g}$ in the definition of $\phi_0$ is absorbed in the
gravitational dressing of the operators. Otherwise $\phi_0$ varies
with $\hat g$ and we can no longer expect that the perturbations
are (1,1), let alone exactly marginal.}
It just requires
specifying the action at the cutoff, $S(\phi=\phi_0)$.
As in [\polch], we identify it with the unperturbed action
$S_0$ plus the bare matter interaction
($\Delta$ is the bare cosmological constant:)
$$S(\phi=\phi_0)=S_0+{1\over{8\pi}}
\int\Delta+m_B\cos px\ \Leftrightarrow\ \cases{
&$T(\phi_0)=\Delta+m_B\cos px$\cr
&$G_{\mu\nu}(\phi_0)=\delta_{\mu\nu}$\cr}\eqno(A.4)$$

(ii) IR: It has been pointed out${}^{[\sei,\pch]}$ that
operators that
diverge faster than $e^{-Q/2\ \phi}$ as $\phi\rightarrow -\infty$ do not
exist in the Liouville theory (2.1).
This provides the second boundary condition. Given one solution of (A.3)
for $T^{(1)}$ and $h_{\mu\nu}$, the other solutions are obtained by adding
linear combinations of $o(m^2)$ of
the on-shell tachyons and the
two discrete gravitons
$$\cos px\ e^{(p-{\sqrt2})\phi}
,\ \ \ \cos px\ e^{(-p-{\sqrt2})\phi}
,\ \ \partial x^2\ \ \hbox{and}\ \
\partial x^2\ e^{-2{\sqrt2}\phi}.\eqno(A.5)$$
Boundary condition (ii)
means essentially
that the operators with the more negative Liouville dressing
must be dropped.
For a more precise statement, see appendix B.

The discussion has been restricted to the tachyon and the graviton.
Including the discrete operators of subsection
2.3 as interactions corresponds to turning on
higher spin backgrounds in the sigma model,
and the same arguments seem to apply.
That two boundary conditions still suffice to specify a solution
is suggested by the fact that there are only two possible
Liouville dressings for each of the discrete operators of the
$c=1$ model.

Setting the bound $\phi_0=0$, we see that the operators found in
section 2 already satisfy the boundary conditions
(i) and (ii), and are thus the unique marginal perturbations.
\vskip 1.3cm

{}$\underline{\hbox{Appendix B: The Cosmological Constant}}$

Gravitational dressings
in the presence of a cosmological constant $\mu$
can in principle be found as as follows (See [\polch,\pch] for some details):

One includes the cosmological constant in the tachyon of string theory,
replacing, e.g. for the Sine--Gordon model,
the ansatz (A.2) by
$$\eqalign{
&T(x,\phi)=T_\mu(\phi)+m\cos px\ f_\mu(p,\phi)
+m^2 T^{(1)}_\mu(x,\phi)
+..\cr
&G_{\alpha\beta}(x,\phi)
=\delta_{\alpha\beta}^\mu(\phi)+m^2 h_{\alpha\beta}^\mu(x,\phi)
+..}\eqno(B.1)$$
where $T_\mu$ is the cosmological constant and
$f_\mu$, ${T}_\mu^{(1)}$
and ${h}^\mu$ are the modified dressings, exact in $\mu$
order by order in $m$.
\footnote*{Although we cannot expand in
$\mu$, for $\mu>0$ we can expand in $m$.}
Let us assume that $m\ll\mu$ but both are small.

First, one must find $T_\mu$ and $\delta^\mu$ exactly.
$T_\mu$
has the form of a kink centered at a free parameter $\bar\phi$,
related to $\mu$ by $\mu=e^{{\sqrt2}\bar\phi}$ and to the bare
cosmological constant $\Delta$ by $\Delta\propto\bar\phi
e^{{\sqrt2}\bar\phi}$:
${}^{[\polch]}$
$$T_\mu(\phi)=T_0(\phi-\bar\phi),\ \ \ T_0(\phi)=\cases{
1 &for $\phi\rightarrow-\infty$\cr
\propto\ \phi e^{-{\sqrt2}\phi} &for $\phi\rightarrow\infty$\cr
 }\eqno(B.2)$$
(B.2) satisfies
the boundary conditions (i), $T_\mu(0)=
\Delta$
and (ii), $T_\mu$ is regular as $\phi\rightarrow-\infty$
($T\rightarrow1$). $\delta^\mu$
differs from $\delta$ by the backreaction of the tachyon $T_\mu$ on
the metric. Like $T_\mu$, this difference
will decay exponentially in the UV.

Next, one must find the dressings
$f_\mu$, ${T}_\mu^{(1)}$
and ${h}^\mu$
by solving the string equations of motion (A.3) order by order in $m$.
E.g., the tachyon equation of motion,
linearized around the
background $T_\mu$, determines
$f_\mu$:${}^{[\polch]}$
$$\{\partial_\phi^2+2\stwo
\partial_\phi+2-p^2-4T_\mu\}f_\mu=0.
\eqno(B.3)$$
Since $T_\mu$ and
$\delta^\mu$ are very small in the UV
$(\bar\phi\ll\phi<0)$,
the equations of
motion for $f_\mu$, ${T}_\mu^{(1)}$
and ${h}^\mu$
are the same as for $\mu=0$ in this regime
and the only role of the cosmological constant is
to set the second boundary condition (ii) of appendix A.
E.g., the solutions
of (B.3) in the UV are${}^{[\polch]}$
$$c_1e^{(-p-\stwo)\phi}
+ c_2e^{(+p-\stwo)\phi}
\ \propto\  e^{-\stwo\phi}\sinh(p\phi-\Theta).$$
In the IR region $\phi\ll\bar\phi$, where $T_\mu\sim$ constant,
the solution of (B.3)
that is regular
at $\phi\rightarrow -\infty$ grows exponentially.
The other, divergent solution
does not exist as an operator.
To match the solutions for
$\phi<\bar\phi$ and $\phi>\bar\phi$, one needs roughly
$\Theta\sim p\bar\phi$. In the UV, $\phi-\bar\phi$ is large,
of order $\vert\log a^2\vert$. So unless
$p$ is close to zero, $f_\mu$ is just
$e^{(p-\stwo)\phi}$ there.
Boundary condition (ii) then simply means dropping the term with
the second Liouville dressing, as without cosmological constant.
At $o(m^2)$, the same arguments can be repeated for
${T}_\mu^{(1)}$
and ${h}^\mu$.

Next, let us discuss OPE's in the presence of the cosmological constant.
In free field theory, the OPE of two operators
with Liouville momenta $\alpha,
\beta$ would produce an operator with Liouville momentum
$\alpha+\beta$. But in Liouville theory
momentum is not conserved because of the
exponential potential. Also, if $\alpha+\beta<-Q/2$, the operator
$e^{(\alpha+\beta)\phi}$ does not exist.
Instead, new primary fields $V_\sigma=e^{-Q/2\ \phi}\sin(\sigma\phi+\Theta)$
will be produced, with some weight $f(\sigma)$ and
less singular coefficients:${}^{[\pch]}$
$$\eqalign{
e^{\alpha\phi(r)}\ e^{\beta\phi(0)}\ \sim\int^\infty_0{d\sigma}
&\vert r\vert^{-2\alpha\beta+(\alpha+\beta+Q/2)^2+\sigma^2}
f(\sigma)\ V_\sigma\cr \hbox{instead of\hskip 2cm}
&\vert r\vert^{-2\alpha\beta}\ e^{(\alpha+\beta)\phi}.}\eqno(B.4)$$
For the Sine--Gordon model, this modification of the OPE's seems to cure
the problem of new $\vert r\vert^{-2}$
singularities that would naively appear in (2.7)
below $p={1\over2}\stwo$. They would give rise to unwanted
counterterms like $\cos 2px$ at $p={1\over4}\stwo$ and
$\partial x^2 \cos 2px$ at $p=0$. The
modified OPE of $\cos px\ e^{\epsilon\phi}$ with itself produces
$$
\int {{d\sigma}\over{2\pi}}f(\sigma)
\vert r\vert^{-2+4p^2+\sigma^2}\cos 2px\
V_\sigma(\phi)\ +...
\eqno(B.5)$$
Except for the (negligible) case $p=\sigma=0$,
all singularities are milder than quadratic.

\vskip 2in
\par\penalty-400\vskip\chapterskip\spacecheck\referenceminspace
   \ifreferenceopen \Closeout\referencewrite \referenceopenfalse \fi
   \line{\fourteenrm\hfil REFERENCES\hfil}\vskip\headskip
   \input referenc.txa
   
\vfill\eject
\centerline{Figures}
\vskip 3.5in
\centerline{Fig. 1}
\centerline{KT--transition with gravity at $o(m^2)$; arrows point towards IR}
\vskip 3.5in
\centerline{Fig. 2}
\centerline{Sine--Gordon model with gravity at $o(m^2)$}

\end